\documentclass{raa}

\usepackage{graphicx,times}
\usepackage{natbib}
\usepackage{amssymb,amsmath}
\bibpunct{(}{)}{;}{a}{}{,}

\usepackage{textcomp}
\allowdisplaybreaks[4]
\usepackage{dsfont}

\usepackage[a4paper=true,dvipdfm=true,pagebackref=true]{hyperref}
\hypersetup{pdftitle = The title of my PDF, pdfauthor = My name, pdfsubject= The subject, pdfkeywords = keyword1 keyword2 keyword3} 
\hypersetup{colorlinks = true, linkcolor = green, anchorcolor = red, citecolor = blue, filecolor = red, pagecolor = red, urlcolor = red}

\newcommand{\dm}{\ensuremath{D_{\mu\mu}}}
\newcommand{\const}{\ensuremath{\text{const}}}
\newcommand{\abs}[1]{\ensuremath{\left\lvert #1\right\rvert}}

\newcommand{\eps}{\ensuremath{\varepsilon}}

\renewcommand{\be}{\begin{equation}}
\renewcommand{\ee}{\end{equation}}
\newcommand{\bs}{\begin{subequations}}
\newcommand{\es}{\end{subequations}}

\newcommand{\R}{\ensuremath{\mathds{R}}}

\newcommand{\pa}{\ensuremath{_\parallel}}
\newcommand{\se}{\ensuremath{_\perp}}

\newcommand{\Om}{\ensuremath{\varOmega}}
\newcommand{\Th}{\ensuremath{\varTheta}}

\newcommand{\f}[1]{\ensuremath{\boldsymbol{#1}}}

\newcommand{\pd}[2][]{\ensuremath{\frac{\partial #1}{\partial #2}}}
\newcommand{\dd}[2][]{\ensuremath{\frac{\mathrm{d} #1}{\mathrm{d} #2}}}
\newcommand{\df}{\ensuremath{\mathrm{d}}}

\begin{document}

\title{Application of the three-dimensional telegraph equation to cosmic-ray transport}

\volnopage{ {\bf 2012} Vol.\ {\bf X} No. {\bf XX}, 000--000}
\setcounter{page}{1}

\author{R.\,C. Tautz\inst{1} and I. Lerche\inst{2}}
\institute{Zentrum f\"ur Astronomie und Astrophysik, Technische Universit\"at Berlin, Hardenbergstra\ss{}e 36, D-10623 Berlin, Germany {\it robert.c.tautz@gmail.com}\\
\and Institut f\"ur Geowissenschaften, Naturwissenschaftliche Fakult\"at III, Martin-Luther-Universit\"at Halle, D-06099 Halle, Germany {\it lercheian@yahoo.com}\\
\vs \no
{\small Received May 9, 2016; accepted June 27, 2016}
}

\abstract{An analytical solution to the the three-dimensional telegraph equation is presented. This equation has recently received some attention but so far the treatment has been one-dimensional. By using the structural similarity to the Klein-Gordon equation, the telegraph equation can be solved in closed form. Illustrative examples are used to discuss the qualitative differences to the diffusion solution. The comparison with a numerical test-particle simulation reveals that some features of an intensity profile can be better explained using the telegraph approach.
}

\authorrunning{R.\,C. Tautz \& I. Lerche}            
\titlerunning{Three-dimensional telegraph equation}  
\maketitle

\section{Introduction}

In discussing the early phase of turbulent transport, recently the telegraph equation \citep[originally due to][]{hea89:ele} has been invoked as a possible alternative to the diffusion equation \citep[e.\,g.,][]{lit13:tel,lit13:num,lit15:tel}. The reason is that, for early times, the particles move ballistically until they become scattered for the first time. In the running diffusion coefficient---as determined for instance by numerically tracing the mean-square displacement of a particle ensemble---this behavior is reflected in a linear increase in agreement with a ballistic motion. A second problem is the finite propagation speed of the particles, which (in the limiting case of magnetostatic turbulence) have a fixed energy and thus cannot have a finite probability to fill space at large distances from the source, as suggested by the solution of the usual diffusion equation.

The telegraph equation, in contrast, has the potential to differ between early ballistic motion and later diffusive transport. The reason is an additional time scale $\tau$, which separates the influence of the first-order time derivative, responsible for the diffusion, and the second-order time derivate, which causes a wave-like behavior and is therefore used to describe the propagation of a pulse along a wire. At least for early times, this behavior seems to be more in agreement with the finite propagation speed of charged particles at a given energy.

Applications of the telegraph approach include the transport of solar energetic particles, which, depending on their energy (or rigidity), indeed sometimes arrive during their ballistic phase \citep{eff14:te1,abl16:sol}. Some aspects regarding satellite observations of such events are still not understood and various explanations have been invoked, ranging from extended source regions to enhanced cross-field diffusion \citep[e.\,g.,][]{Dresing2012_longspread,lai13:cro}.

A potential problem in applying the standard telegraph equation to the turbulent motion of charged particles is that there is no inherent one-dimensionality. On the contrary, due to the preferred direction of the mean magnetic field, the three-dimensional (or at the very least, two-dimensional) character of the problem is emphasized. This requirement is reflected in the fact that the diffusion coefficients for perpendicular transport are smaller than the parallel diffusion coefficient by at least one order of magnitude. In addition, for the diffusion equation the dimensionality is reflected in the index of the power law in time.

Therefore, here we present an analytical solution to the three-dimensional telegraph equation. Such a fully three-dimensional treatment is necessary in order to shed light on the question of whether or not the telegraph equation can provide a better description of turbulent particle propagation at early times. More recently, a criticism was raised by \citet{mal15:tel}, who stated that no systematic derivation of the telegraph equation was given in application to charged particle transport. Moreover, the norm of the solution is not conserved, which they claimed would render the model invalid. Here we show that there is a natural explanation for this behavior.

This article is organized as follows. In Sec.~\ref{tech}, a brief motivation for the use of the telegraph equation is given and the solution is derived. In Sec.~\ref{illus}, examples are shown in comparison to the solution of the diffusion equation and the implications are discussed. Sec.~\ref{summ} provides a discussion of the results and a brief outlook.

\section{Technical Development}\label{tech}

A systematic derivation of the telegraph equation in application to charged particle transport has been given by \citet{lit13:tel}. In contrast to their treatment, here we neglect adiabatic focusing, which is expressed through a focusing length $L=-B/(dB/dz)$. Indeed, \citet{lit13:tel} point out that their derivation is valid only for the case of weak focusing, so that we can formally set $B=\const$, which results in $L=\infty$.

The telegraph time scale, $\tau$, is then given through
\be
\tau=\frac{3v}{8\lambda\pa}\int_{-1}^1\df\mu\left(\int_0^\mu\df\mu'\;\frac{1-{\mu'}^2}{\dm(\mu')}\right)^2
\ee
with the Fokker-Planck coefficient $\dm(\mu)$ describing random changes in the particle's pitch angle \citep[see][for overviews]{rs:rays,sha09:nli}. The parallel mean-free path is given by
\be
\lambda\pa=\frac{3v}{8}\int_{-1}^1\df\mu\;\frac{\left(1-\mu^2\right)^2}{\dm(\mu)},
\ee
which also describes the transition from the Fokker-Planck equation to a diffusion equation, because of the relation $\lambda\pa=(v/3)\kappa\pa$ with $\kappa\pa$ the parallel diffusion coefficient.

By assuming the quasi-linear approximation \citep{jok66:qlt}, the pitch-angle scattering coefficient can be expressed as $\dm\propto\abs\mu^{q-1}(1-\mu^2)$, which leads to a simplified telegraph time scale \citep{lit13:tel}
\be
\tau=\frac{\lambda\pa}{v}\,\frac{\left(4-q\right)^2}{3\left(5-2q\right)}.
\ee

For gyrotropic magnetic turbulence, the diffusion coefficients for perpendicular transport are equal, which however is not a necessary constraint for the following analysis. The three-dimensional telegraph equation then reads
\be\label{eq:tel}
\pd[f]t+\tau\,\pd[^2f]{t^2}=\kappa\pa\,\pd[^2f]{z^2}+\kappa\se\left(\pd[^2f]{x^2}+\pd[^2f]{y^2}\right)
\ee
which, in the case of $\tau\to0$, reduces to the usual three-dimensional diffusion equation.

\subsection{Derivation of the solution}

\noindent
Set $x=\sqrt{\kappa\se}\,\xi$, $y=\sqrt{\kappa\se}\,\eta$, and $z=\sqrt{\kappa\pa}\,\zeta$ to obtain
\be\label{eq:tmp1}
\pd[f]t+\tau\,\pd[^2f]{t^2}=\pd[^2f]{\zeta^2}+\pd[^2f]{\xi^2}+\pd[^2f]{\eta^2}=:\nabla^2f.
\ee

To proceed, multiply Eq.~\eqref{eq:tmp1} with $e^{t/2\tau}$ and set $g=e^{t/2\tau}f$ when
\be
e^{t/2\tau}\pd[f]t+\tau\,e^{t/2\tau}\pd[^2f]{t^2}=\tau\,\pd[^2g]{t^2}-\frac{g}{4\tau}.
\ee
By replacing $f$ with $g$, this transformation results in a Klein-Gordon equation \citep{cha66:klg} of the form
\be\label{eq:kg}
\nabla^2g-\tau\,\pd[^2g]{t^2}+\frac{g}{4\tau}=0.
\ee

Fortunately, the solution to Eq.~\eqref{eq:kg} is known {for the case of a point source of particles at a single point in time, i.\,e., $f(t=0)=\delta(t)$.} Compare with Eq.~(7.3.33) of \citet{mor53:th1} so that their parameters become $c^2=1/\tau$ and $\kappa^2=-1/4\tau$. The solution is then
\be
g(\varrho,t)=\frac{\delta\!\left(t-\varrho/c\right)}{\varrho}\,J_0\left(\kappa c\sqrt{t^2-(\varrho/c)^2}\right)-\frac{\kappa}{\sqrt{t^2-(\varrho/c)^2}}\,J_1\left(\kappa c\sqrt{t^2-(\varrho/c)^2}\right)
\ee
in $t>\varrho/c$, where $\delta(\cdot)$ is the Dirac delta distribution and where the normalized radial coordinate is expressed as
\be\label{eq:rho}
\varrho=\sqrt{\xi^2+\eta^2+\zeta^2}=\sqrt{\frac{x^2+y^2}{\kappa\se}+\frac{z^2}{\kappa\pa}}.
\ee

For the Bessel function of the first kind, $J_\nu(\cdot)$, use the property $J_1(i\alpha)=i\,I_1(\alpha)$ for $\alpha\in\R$, where $I_\nu(\cdot)$ is the modified Bessel function of the first kind. The solution to the three-dimensional telegraph equation is then
\be\label{eq:sol}
f(\varrho,t)=C\,e^{-t/2\tau}\Biggl[\frac{\delta\!\left(t-\varrho\sqrt\tau\right)}{\varrho}\,I_0\left(\frac{1}{2}\sqrt{\frac{t^2}{\tau^2}-\frac{\varrho^2}{\tau}}\right)+\frac{\Th\left(t/\sqrt\tau-\varrho\right)}{\displaystyle2\tau^{3/2}\sqrt{\frac{t^2}{\tau^2}-\frac{\varrho^2}{\tau}}}\;I_1\left(\frac{1}{2}\sqrt{\frac{t^2}{\tau^2}-\frac{\varrho^2}{\tau}}\right)\Biggr], 
\ee
where $\Th(\cdot)$ is the Heaviside step function and $C$ is a constant. Accordingly, the second part of the solution is non-zero only for $\varrho<t/\sqrt\tau$, which reflects the fact that all particles require a finite time to arrive at regions more remote from the source. The first part of the solution describes a narrow beam with diminishing amplitude as both time and distance from the source increase.

\begin{figure}[tb]
\centering
\includegraphics[width=0.55\linewidth]{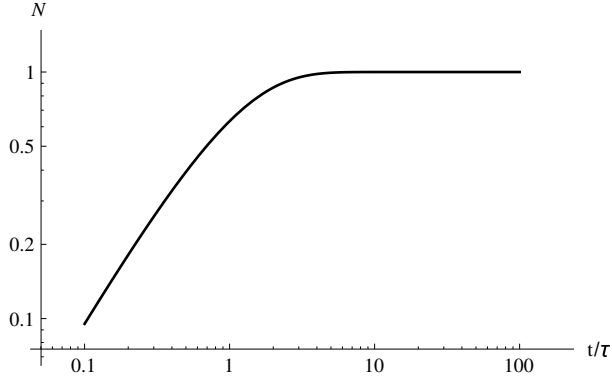}
\caption{Normalization given by Eq.~\eqref{eq:nrm} for the solution to the three-dimensional telegraph equation, Eq.~\eqref{eq:sol}, as a function of the dimensionless time $t/\tau$.}
\label{ab:F3}
\end{figure}

{A referee has noted that the solution to the three-dimensional telegraph equation is also applicable to the two-dimensional situation if one sets $\kappa\pa=\infty$ and is also applicable to the one-dimensional situation if one sets $\kappa\se=\infty$, yielding $\varrho=z/\sqrt{\kappa\pa}$. In this latter case, a comparison with the solution of the one-dimensional telegraph equation presented in \citet[their Eq.~(29)]{lit15:tel} shows largely agreement. However, the (exponentially suppressed) delta distributions \citep[see also][their Eqs.~(18) and (19)]{eff14:te1} are not present \citep[but cf.][]{lit13:tel}.}

In the limit $\tau\to0$, the solution must reduce to that of the three-dimensional diffusion equation, which fact allows one to determine the constant factor $C$. Use
\be\label{eq:diff}
f(\varrho,t)\xrightarrow{\tau\to0}\frac{C}{2\sqrt\pi t^{3/2}}\,\exp\left(-\frac{\varrho^2}{4t}\right)\stackrel{!}{=}\frac{1}{\left(4\pi t\right)^{3/2}\kappa\se\sqrt{\kappa\pa}}\,\exp\left(-\frac{\varrho^2}{4t}\right)
\ee 
so that $C^{-1}=4\pi\kappa\se\sqrt{\kappa\pa}$.

Note that the diffusion equation has particles occurring at \emph{all} values of $\varrho$ for any time. This behavior is unphysical because it implies particles have infinite speed to achieve such a behavior. The telegraph equation corrects for this problem by \emph{not} allowing all values of $\varrho$ to have particles at any time. Instead the particles are confined to the domain $\varrho<ct$ which removes the problem associated with the diffusion equation.

Another important observation is that Eq.~\eqref{eq:sol} is not the only solution to the three-dimensional telegraph equation. An example for such a solution is presented in Appendix~\ref{alternative}. Unfortunately, however, that solution does not fulfill the condition for the conservation of the particle number, either.

\subsection{Normalization}\label{norm}

Unlike the solutions of the diffusion equation, which have constant norms, the norm of the telegraph equation is time-dependent as will now be shown. Integrate over $\df^3r$ which, for an isotropic solution (in $\varrho$), means that
\be
\df^3r=(\kappa\se\sqrt{\kappa\pa})^{-1}\,\df^3\varrho=C^{-1}\df\varrho\,\varrho^2.
\ee
For the first part of the solution involving the Dirac delta distribution, one employs
\be
\int_0^\infty\df\varrho\;\varrho\,\delta\!\left(t-\varrho\sqrt\tau\right)I_0\left(\frac{1}{2}\sqrt{\frac{t^2}{\tau^2}-\frac{\varrho^2}{\tau}}\right)=\frac{t}{\tau}.
\ee

\begin{figure}[tb]
\centering
\includegraphics[width=0.55\linewidth]{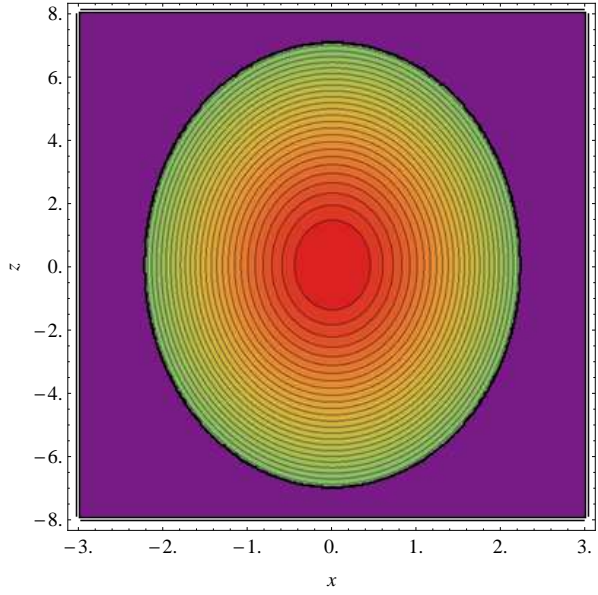}
\caption{(Color online) Contour lines of the solution to the three-dimensional telegraph equation, Eq.~\eqref{eq:sol}, in the $x$-$z$ plane for a fixed time $t=1$. The ratio of the diffusion coefficients entering the normalized radial coordinate $\varrho$ according to Eq.~\eqref{eq:rho} is chosen to be $\kappa\pa/\kappa\se=10$. Note the different scaling of the two axes.}
\label{ab:Tel3D}
\end{figure}

\begin{figure*}[tb]
\centering
\includegraphics[width=0.49\linewidth]{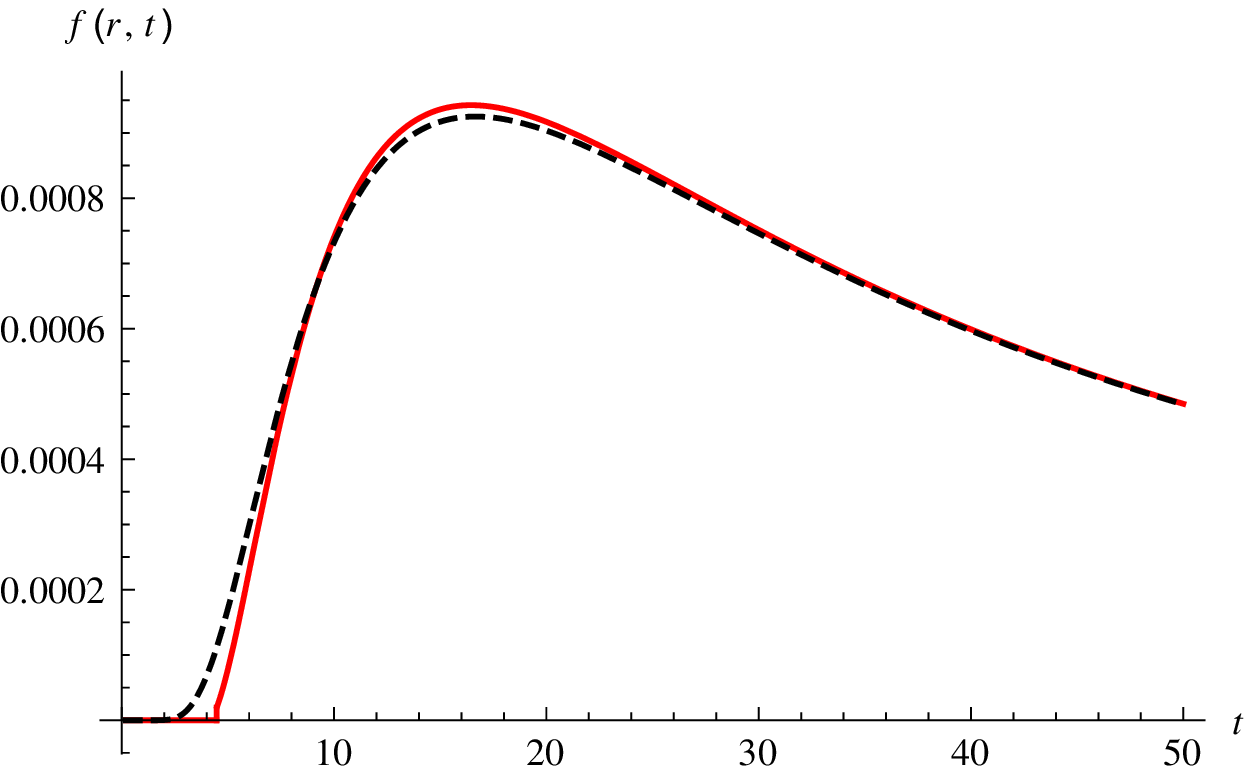}
\includegraphics[width=0.49\linewidth]{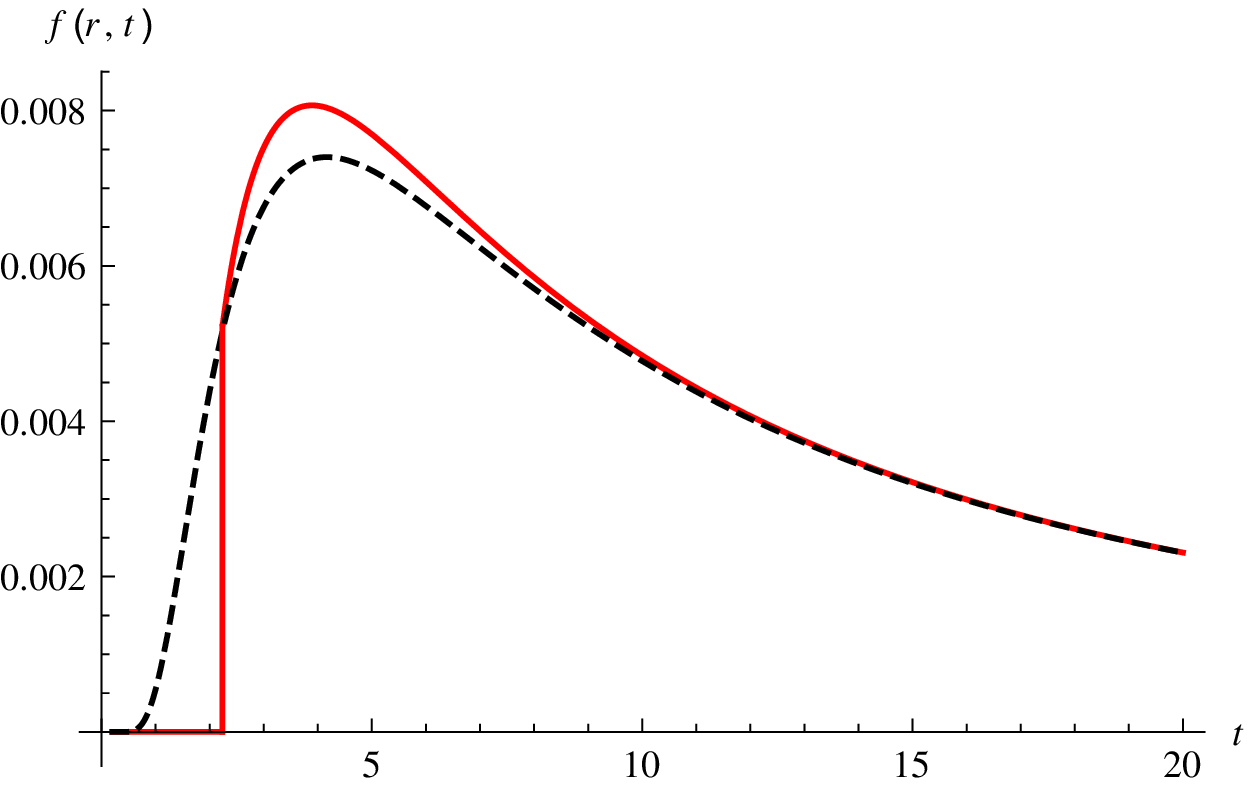}\\[2pt]
\includegraphics[width=0.49\linewidth]{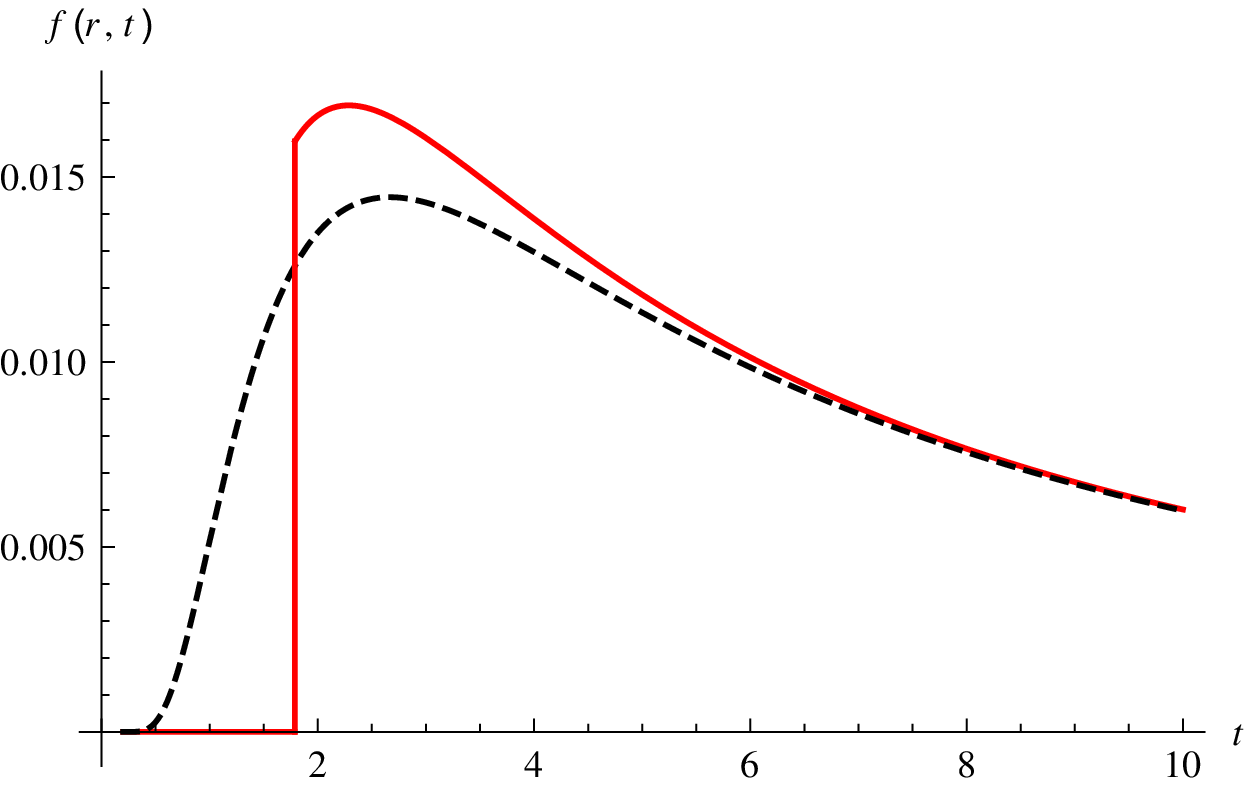}
\includegraphics[width=0.49\linewidth]{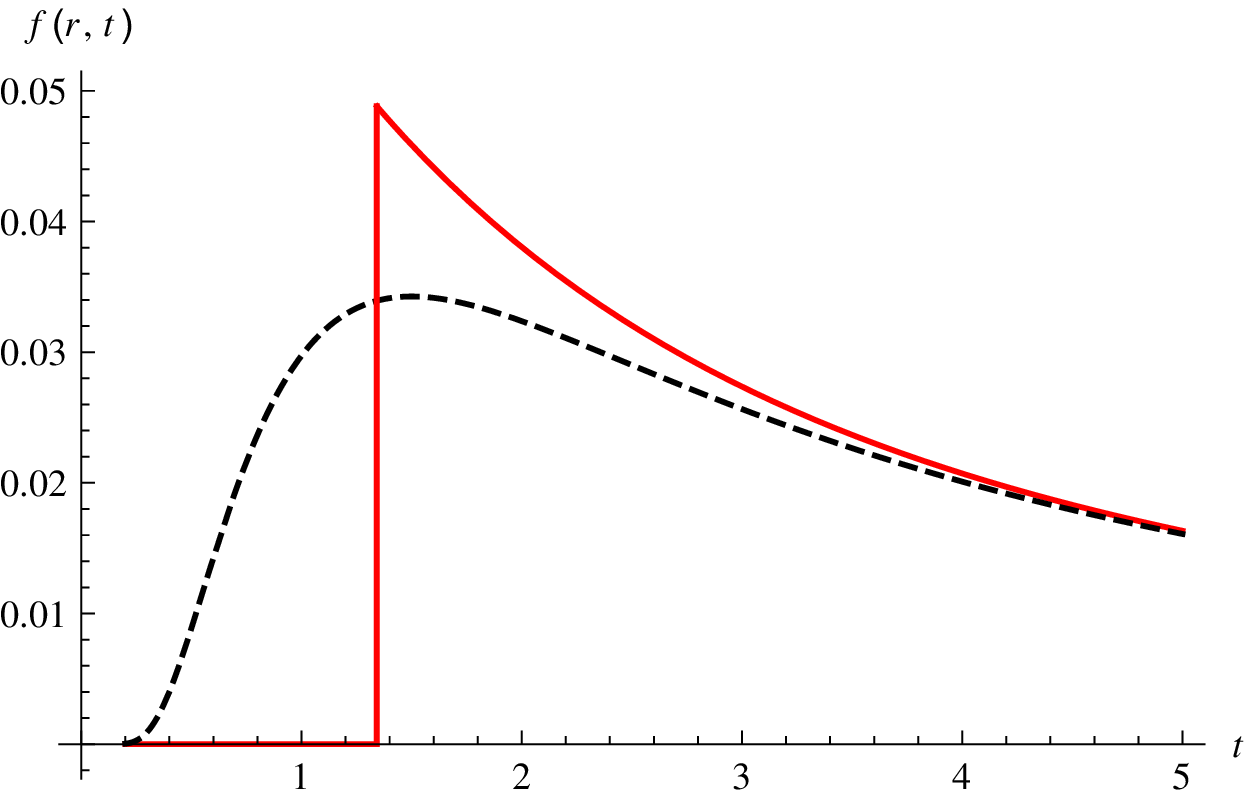}
\caption{(Color online) Solution to the three-dimensional telegraph equation (red solid line), Eq.~\eqref{eq:sol}, in comparison to the diffusion solution (black dashed line) given in Eq.~\eqref{eq:diff}, both as a function of time. The distance from the source increases from $\varrho=10$ (upper left panel) through to $\varrho=5$ (upper right) and $\varrho=4$ (lower left) to $\varrho=3$ (lower right panel) with $\tau=0.2$ in all cases. Here, time and space coordinates are given in arbitrary units. In addition, part of the solution involving the Dirac delta distribution is skipped for clarity.}
\label{ab:F1234}
\end{figure*}

The second part is slightly more involved. Use the recurrence relation of the modified Bessel function,
\be
I_1\left(\frac{1}{2}\sqrt{\frac{t^2}{\tau^2}-\frac{\varrho^2}{\tau}}\right)=-\frac{2\tau}{\varrho}\sqrt{\frac{t^2}{\tau^2}-\frac{\varrho^2}{\tau}}\dd\varrho\,I_0\left(\frac{1}{2}\sqrt{\frac{t^2}{\tau^2}-\frac{\varrho^2}{\tau}}\right).
\ee
Integration by parts then yields
\bs
\begin{align}
&\int_0^{t/\sqrt\tau}\frac{\df\varrho\;\varrho^2}{\sqrt{\displaystyle\frac{t^2}{\tau^2}-\frac{\varrho^2}{\tau}}}\,I_1\left(\frac{1}{2}\sqrt{\frac{t^2}{\tau^2}-\frac{\varrho^2}{\tau}}\right)\nonumber\\
=\;&-2\tau\int_0^{t/\sqrt\tau}\df\varrho\;\varrho\,\dd\varrho\,I_0\left(\frac{1}{2}\sqrt{\frac{t^2}{\tau^2}-\frac{\varrho^2}{\tau}}\right)\\
=\;&-2t\sqrt\tau+2\tau\int_0^{t/\sqrt\tau}\df\varrho\;I_0\left(\frac{1}{2}\sqrt{\frac{t^2}{\tau^2}-\frac{\varrho^2}{\tau}}\right).
\end{align}
\es
To proceed, set $\varrho=(t/\sqrt\tau)\sin\phi$ and use formula~(3) in Sec.~6.681 of \citet{gr:int} when
\be
\int_0^{\pi/2}\df\phi\,\cos\phi\,I_0\left(z\cos\phi\right)=\frac{\pi}{2}\,I_{-\frac12}\left(\frac{z}{2}\right)I_{\frac12}\left(\frac{z}{2}\right)
\ee
and employ the connection of the Bessel function to the hyperbolic sine so that
\be
\frac{\pi}{2}\,\frac{t}{\tau}\,e^{-t/2\tau}I_{-\frac12}\left(\frac{t}{4\tau}\right)I_{\frac12}\left(\frac{t}{4\tau}\right)=1-e^{-t/\tau}.
\ee

\begin{figure}[tb]
\centering
\includegraphics[width=0.55\linewidth]{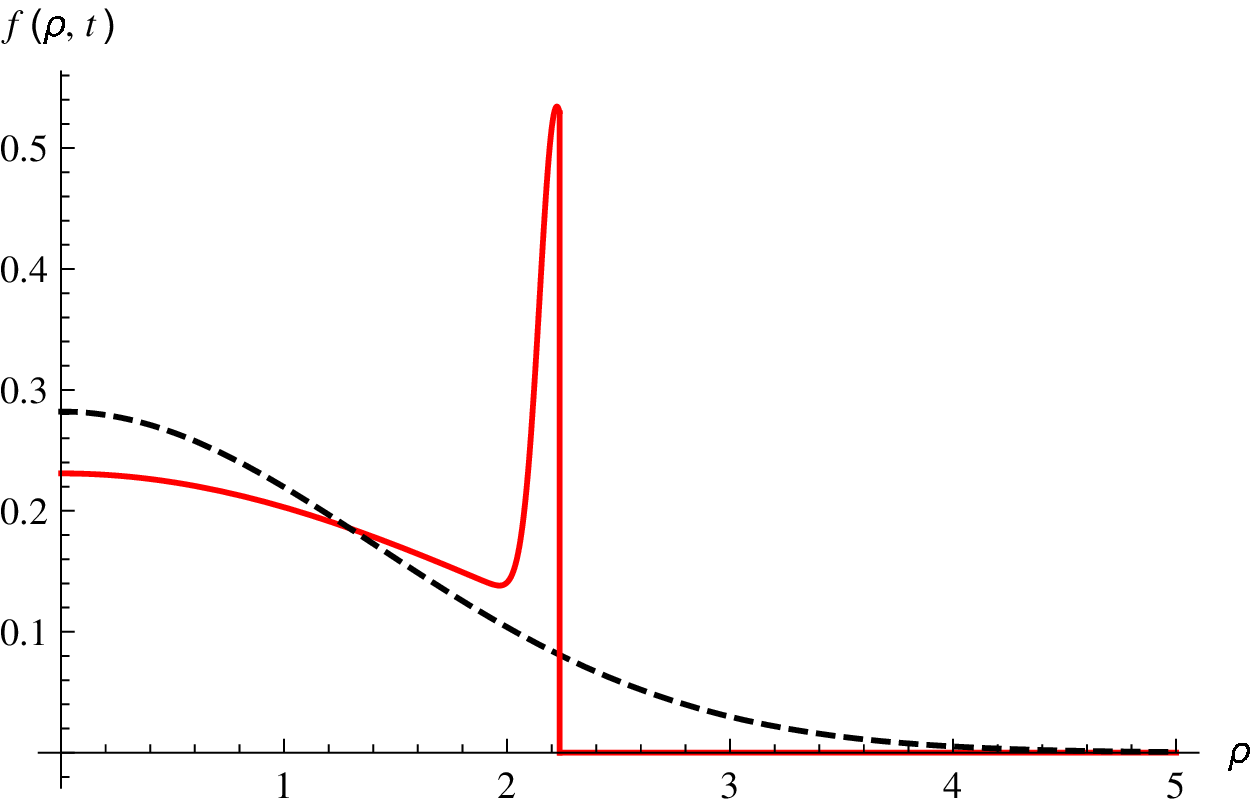}\\[2pt]
\includegraphics[width=0.55\linewidth]{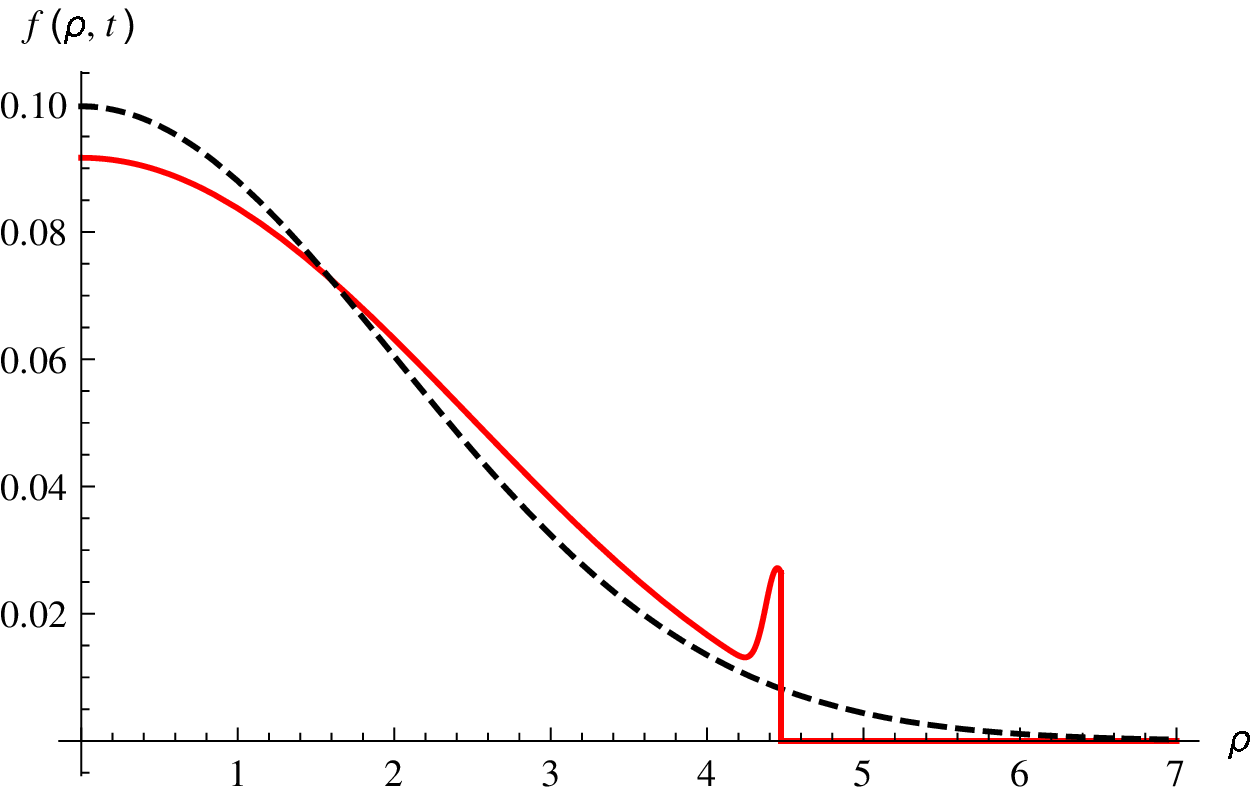}
\caption{(Color online) Solution to the three-dimensional telegraph equation (red solid line), Eq.~\eqref{eq:sol}, in comparison to the diffusion solution (black dashed line) given in Eq.~\eqref{eq:diff}, both as a function of the normalized radial coordinate, $\varrho$. The time at which the solution is taken increases from the upper ($t=1$) to the lower ($t=2$) panel by a factor of two with $\tau=0.2$ in both cases. Note that, for illustration purposes, the Dirac delta peak is approximated by half of a narrow Gaussian.}
\label{ab:G12}
\end{figure}

Collecting terms, the result for the norm of the solution function is
\be\label{eq:nrm}
N\equiv\int\df^3r\;f(r,t)=1-e^{-t/\tau},
\ee
which corresponds to the number of particles. Accordingly, the particle flux is
\be\label{eq:Ndot}
\dot N=\frac{e^{-t/\tau}}{\tau},
\ee
which systematically decreases as $t$ increases from $t=0$ and vanishes for $t\gg\tau$. A convincing interpretation of this change is currently elusive; however, it could be interpreted as the number of particles participating in the diffusion process. For early times, this number increases until, eventually, all particles are transported diffusively and $N$ remains constant.

{The problem of normalization is also evident by considering the flux of cosmic rays,}
\be
\f S=\left(\kappa\se\,\pd[f]x,\kappa\se\,\pd[f]y,\kappa\pa\,\pd[f]z\right).
\ee
{By integrating Eq.~\eqref{eq:tel} over the entire space, one obtains the following result}
\be
\nabla\cdot\f S=\pd[f]t+\tau\,\pd[^2f]{t^2},
\ee
{which clearly shows that the usual continuity equation is modified by a term involving the second time derivative of the distribution function.}

An alternative viewpoint is to regard $f$ as the probability density function so that $(\df/\df t)\int f\;\df^3p$ represents the probability change of a particle being transported either ballistically or diffusively. At small values of $t\ll\tau$ the probability change is dominated by the ballistic motion whereas at large values of $t\gg\tau$ diffusive scattering dominates. The probability change is thus represented by $\dot N$ given in Eq.~\eqref{eq:Ndot}.

\section{Illustrative examples}\label{illus}

The solution to the three-dimensional telegraph equation, Eq.~\eqref{eq:sol}, is illustrated in Fig.~\ref{ab:Tel3D} regarding the behavior with respect to the directions parallel and perpendicular to the mean magnetic field. From Eq.~\eqref{eq:rho} one notes that the solution is an ellipsoid with numerical eccentricity
\be
\eps=\sqrt{1-\left(\frac{\kappa\se}{\kappa\pa}\right)^2}.
\ee

\begin{figure}[tb]
\centering
\includegraphics[width=0.55\linewidth]{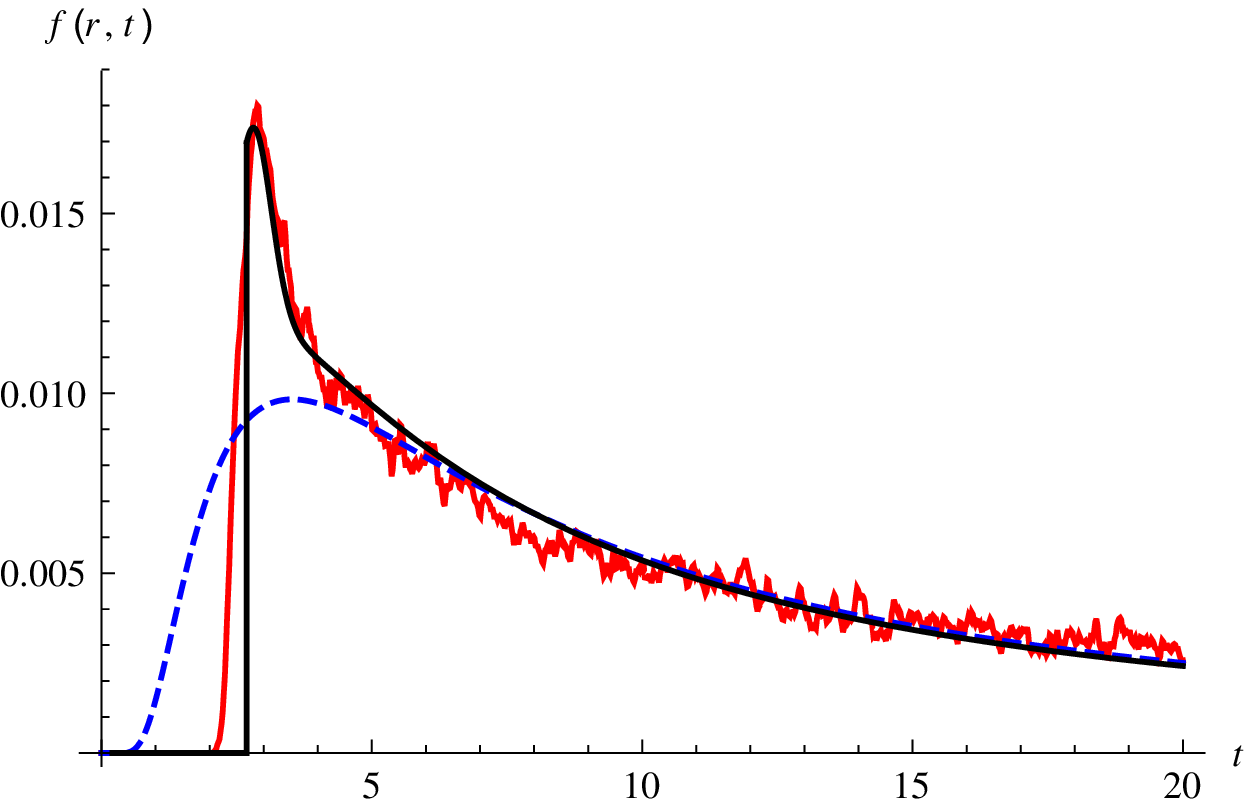}\\[2pt]
\includegraphics[width=0.55\linewidth]{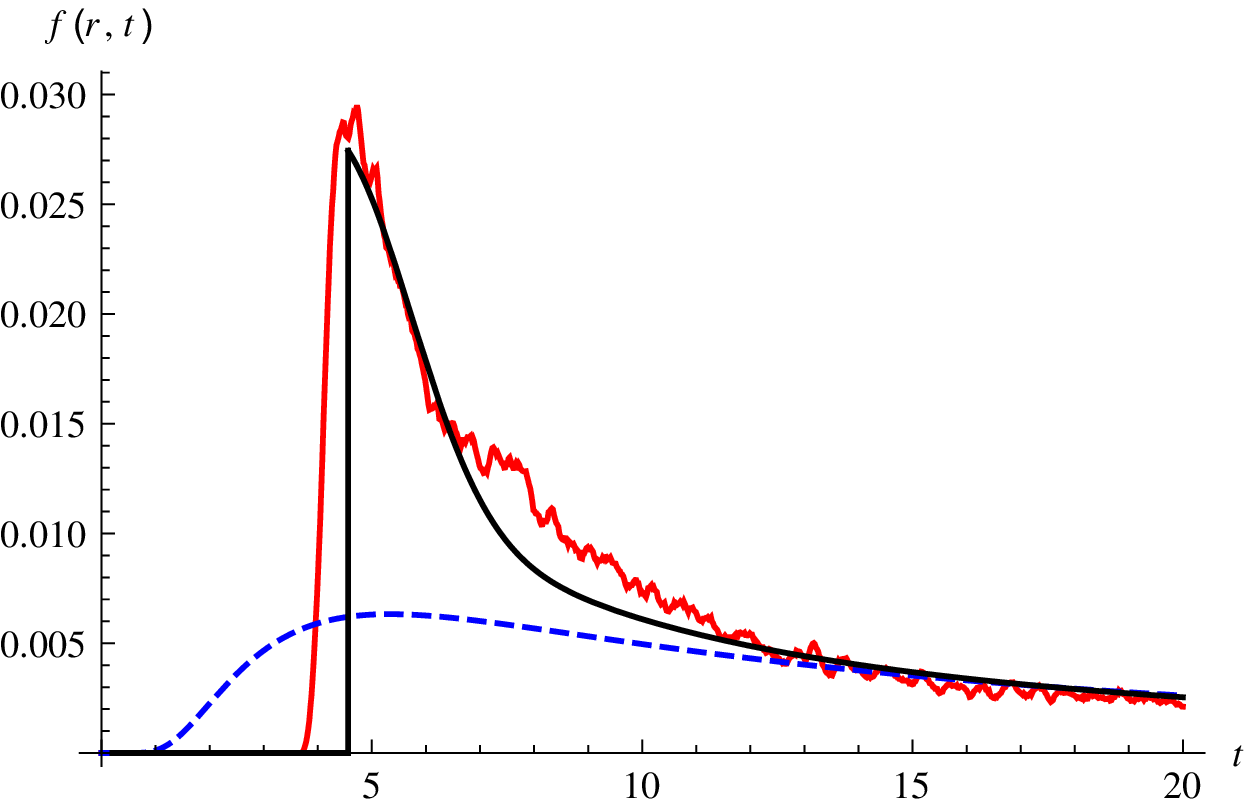}\\[2pt]
\caption{(Color online) Comparison of intensity profiles obtained in a test-particle simulation (red line) together with the solutions to the telegraph (black line) and diffusion (blue dashed line) equations. In the upper panel, the distance $\varrho$ is ten times larger than in the lower panel. Note that both the intensity and the time are given in arbitrary units.}
\label{ab:comp}
\end{figure}

In Fig.~\ref{ab:F1234}, the solutions of the three-dimensional telegraph and diffusion equations are shown in comparison as functions of time at a fixed position in space. This figure illustrates that, particularly for the early transport phase, major differences arise due to the use of the telegraph equation. At later stages of the transport, in contrast, the two solutions become indistinguishable. In particular, one can see the competition of diffusion and ballistic transport so that the initial rise of the usual diffusive solution is not seen if the target is too far away. The overall shape of the solution is thus changed both qualitatively and quantitatively.

Snapshots at fixed times of the total solution including the Dirac delta distribution are shown in Fig.~\ref{ab:G12}. Here, one notes that, due to the early ballistic particle motion, the particle density at the source (at $\varrho=0$) is more rapidly reduced in comparison to the diffusion solution. In addition, the peak describing particles that have not yet been scattered is diminished as time increases. This fact is in agreement with the finding that the number of diffusive particles increases until, for $t\gg\tau$, it becomes constant (i.\,e., all particles moving diffusively).

For comparison purposes, either measurements by spacecraft such as STEREO and ACE or numerical simulation can be used. For instance, test-particle simulations calculate in detail the scattering of an ensemble of particles by a turbulent magnetic field, which allows one to register the number of particles arriving at a given position \citep{tau16:pro}. A detailed description of this Monte-Carlo simulation technique can be found elsewhere \citep[e.\,g.,][and references therein]{tau10:pad,tau13:num}.

In Fig.~\ref{ab:comp}, the solution to the telegraph equation is shown together with an intensity profile obtained from a simulation of particles moving in a magnetic field consisting of a turbulent and a homogeneous (mean) component. The comparison shows that the telegraph solution is able to reproduce both qualitatively and quantitatively the simulated profile, in particular regarding the peak intensity in relation to the late decay phase. This agreement is underlined by the corresponding diffusion solution, for which one clearly sees that: (i) the initial rise begins too early; (ii) the maximum is too broad; and (iii) the peak intensity cannot be reproduced if the normalization is such that agreement is found at late times.

\section{Discussion and Conclusion}\label{summ}

The telegraph equation has recently been revisited as a possible alternative to the diffusion equation, in particular for the transport of energetic charged particle in turbulent magnetic fields such as low-energy cosmic rays in the solar wind. The main reason is that the diffusion equation is not applicable to describe the transport for early times because it yields a non-zero probability density everywhere, which would correspond to an infinite propagation speed. The telegraph equation, in contrast, combines diffusion with a finite propagation speed and so results in a more realistic behavior in particular for the early phase transport.

{The temporal and spatial conditions on the solution to the telegraph equation were taken to be the simplest possible, namely an infinite space traversed by a homogeneous mean magnetic field and a source of particles at a single point in time. The reason for the attempts to use the telegraph equation as a replacement for the diffusion equation is that the diffusion equation under the same temporal and spatial conditions provides a solution with particles at all spatial positions immediately after insertion although the total particle number is conserved. This problem requires that particles travel faster than the speed of light. The telegraph equation overcomes this problem by limiting particle speeds to less than that of light but has the disadvantage that one cannot then conserve particle number. It would seem that a better alternative to both attempts is required.}

However, until now typically only the one-dim\-en\-sion\-al version has been discussed \citep[e.\,g.,][]{lit13:tel,eff14:te1,lit15:tel}. In this article, therefore, a solution to the three-dimensional telegraph equation was derived by exploiting the structural similarity to the Klein-Gordon equation. It was shown that the solution exhibits fundamental differences to both the one-dimensional solution and to the the diffusion solution.

Whether or not the new solution can indeed help one to understand the early ballistic transport phase remains to be seen. Indeed, the comparison to a numerical test-particle simulation has shown that some of the features can be best explained by employing the telegraph profile. However, detailed comparisons with observations and/or numerical simulations and parameter studies are beyond the scope of this article. This subject is currently under active investigation, and results will be presented in due time.


\appendix
\section{An alternative solution}\label{alternative}

Set $x=\sqrt{\kappa\se\tau}\,\xi$, $y=\sqrt{\kappa\se\tau}\,\eta$, $z=\sqrt{\kappa\pa\tau}\,\zeta$, and $t=t_{\text r}/\tau$ to obtain
\be\label{eq:tmp4}
\pd[f]t+\pd[f]{t^2}-\nabla^2f=S(\f r,t)
\ee
with $\f r=[\xi,\eta,\zeta]$ and
\be
\nabla^2f:=\pd[^2f]{\zeta^2}+\pd[^2f]{\xi^2}+\pd[^2f]{\eta^2}.
\ee

To proceed, consider the associated Green's function $\mathcal G(\f r,\f r',t,t')$ which is determined through
\be
\pd[\mathcal G]t+\pd[^2\mathcal G]{t^2}-\nabla^2\mathcal G=\left(2\pi\right)^4\delta(\f r-\f r')\,\delta(t-t').
\ee
The causality of the injected source particles demands that
\be
\mathcal G(\f r,\f r',t,t')=
\begin{cases}
\;G(\f r-\f r',t-t'), & t>t'\\
\;0, & t<t'
\end{cases}
\ee
so that
\be
f(\f r,t)=\int\df^3r'\int_0^t\df t'\;S(\f r',t')G(\f r-\f r',t-t').
\ee

A Fourier transformation in both space and time with dependence $\exp[i(\f k\cdot\f r-\omega t)]$ yields
\be
G(\f k,\omega)\left(-i\omega-\omega^2+k^2\right)=\exp\left(-i\f k\cdot\f r'+i\omega t'\right).
\ee
To facilitate the further analysis, introduce $\Om=\omega+i/2$ so that
\be
\omega^2+i\omega=\Om^2+\tfrac14,
\ee
yielding
\be
G(\f k,\omega)\left(\Om^2+\tfrac14-k^2\right)=\;-\exp\left(-i\f k\cdot\f r'+i\Om t'+\tfrac{1}{2}t'\right).
\ee

In position space, Green's function is thus given by
\be
G(\f r,\f r',t,t')=-\int\df\Om\;\exp\left[-\left(i\Om+\tfrac12\right)\left(t-t'\right)\right]\int\df^3k\;\frac{\exp\left[i\f k\cdot\left(\f r-\f r'\right)\right]}{\Om^2-\left(k^2-\tfrac14\right)}.
\ee
Now choose the coordinate system such that, with $\f\varrho:=\f r-\f r'$, one has $\f k\cdot\f\varrho=k\varrho\mu$ with $\mu=\cos\angle(\f k,\f\varrho)$. Then
\be
\int_{-1}^1\df\mu\;e^{ik\varrho\mu}=\frac{2\sin k\varrho}{k\varrho}.
\ee
The integral over the three-dimensional wavenumber space is thus simplified and one has
\be\label{eq:G}
G=-\frac{4\pi}{\varrho}\int_0^\infty\df k\;k\sin k\varrho\int\df\Om\;\frac{\exp\left[-\left(i\Om+\tfrac12\right)\left(t-t'\right)\right]}{\Om^2-\left(k^2-\tfrac14\right)}.
\ee

\begin{figure}[tb]
\centering
\includegraphics[bb=160 280 500 578,width=0.55\linewidth]{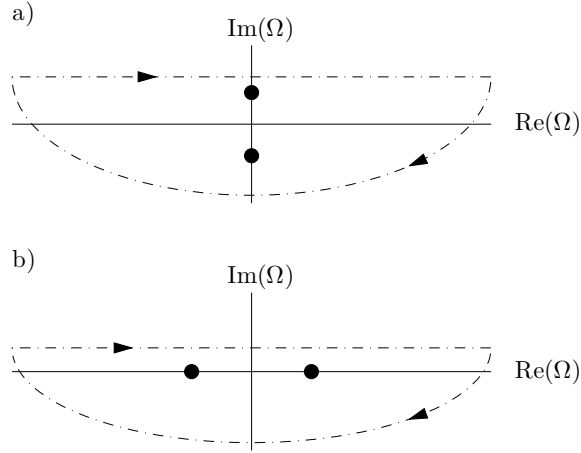}
\caption{Complex contour $\mathcal C$ for the $\Om$ integral in Eq.~\eqref{eq:G}. The poles, as indicated by the black circles, are enclosed by the path of integration, which is closed at infinity in negative mathematical direction of rotation.}
\label{ab:contour}
\end{figure}

Consider the singularities of the integrand in $\Om$ space. Depending on $k^2$ being larger or smaller than $1/4$, there are two real or imaginary poles at
\be
\Om_\pm=\pm
\begin{cases}
\;\sqrt{k^2-\tfrac14}, & k^2>\tfrac14\\
\;i\sqrt{\tfrac14-k^2}, & k^2<\tfrac14
\end{cases}.
\ee
For the $\Om$ integration, the contour must be a path such that $G=0$ for $t<t'$ and $G$ finite and real for $t>t'$. Note that $\exp[-i\Om(t-t')]$ converges in the lower half of the complex plane for $t>t'$. Accordingly, the path of integration consists of a straight line along the real axis and above the poles, and is closed in the lower half plane as depicted in Fig.~\ref{ab:contour}.

For the analytical evaluation of the $\Om$ integral in Eq.~\eqref{eq:G}, split the $k$ integral into the ranges $0\leqslant k\leqslant\tfrac12$ and $\tfrac12\leqslant k<\infty$. Then
\begin{align}
G&=-\frac{4\pi}{\varrho}\,\exp\left[-\frac12\left(t-t'\right)\right]\Biggl[\int_0^{\frac12}\df k\;k\sin k\varrho\int_{\mathcal C}\df\Om\;\frac{\exp\left[-i\Om\left(t-t'\right)\right]}{\Om^2+\left(\tfrac14-k^2\right)}\nonumber\\
&+\int_{\frac12}^\infty\df k\;k\sin k\varrho\int_{\mathcal C}\df\Om\;\frac{\exp\left[-i\Om\left(t-t'\right)\right]}{\Om^2-\left(k^2-\tfrac14\right)}\Biggr].
\end{align}

By making use of the residue theorem, the first $\Om$ integral yields
\be
\int_{\mathcal C}\df\Om\;\frac{\exp\left[-i\Om\left(t\!-\!t'\right)\right]}{\Om^2+\left(\tfrac14-k^2\right)}=-\frac{2\pi\sinh\left[\left(t\!-\!t'\right)\sqrt{\tfrac14-k^2}\right]}{\sqrt{\tfrac14-k^2}},
\ee
where the additional minus on the right-hand side results from the negative winding number of the integration contour $\mathcal C$. Likewise, the second $\Om$ integral gives
\be
\int_{\mathcal C}\df\Om\;\frac{\exp\left[-i\Om\left(t\!-\!t'\right)\right]}{\Om^2-\left(k^2-\tfrac14\right)}=-\frac{2\pi\sin\left[\left(t\!-\!t'\right)\sqrt{k^2-\tfrac14}\right]}{\sqrt{k^2-\tfrac14}}.
\ee

The resulting expression for $G$ is thus
\begin{align}
G&=\frac{8\pi^2}{\varrho}\,\exp\left[-\frac12\left(t-t'\right)\right]\Biggl[\int_0^{\frac12}\frac{\df k\;k\sin k\varrho}{\sqrt{\tfrac14-k^2}}\,\sinh\left[\left(t-t'\right)\sqrt{\tfrac14-k^2}\right]\nonumber\\
&+\int_{\frac12}^\infty\frac{\df k\;k\sin k\varrho}{\sqrt{k^2-\tfrac14}}\sin\left[\left(t-t'\right)\sqrt{k^2-\tfrac14}\right]\Biggr] \label{eq:tmp2}
\end{align}
for $t>t'$ and $G=0$ for $t<t'$.
In the first integral, set $x^2=1-4k^2$ to obtain
\bs\label{eq:i1i2}
\begin{align}
\mathcal I_1&=\int_0^{\frac12}\frac{\df k\;k\sin k\varrho}{\sqrt{\tfrac14-k^2}}\sinh\left[\left(t-t'\right)\sqrt{\tfrac14-k^2}\right]\nonumber\\
&=\frac{1}{2}\int_0^1\df x\,\sin\left(\frac{\varrho}{2}\sqrt{1-x^2}\right)\sinh\left[\left(t-t'\right)x\right]. \label{eq:i1}
\end{align}
Likewise, set $x^2=1+4k^2$ in the second integral of Eq.~\eqref{eq:tmp2} so that
\begin{align}
\mathcal I_2&=\int_{\frac12}^\infty\frac{\df k\;k\sin k\varrho}{\sqrt{k^2-\tfrac14}}\sin\left[\left(t-t'\right)\sqrt{k^2-\tfrac14}\right]\nonumber\\
&=\frac{1}{2}\int_0^\infty\df x\,\sin\left(\frac{\varrho}{2}\sqrt{1+x^2}\right)\sin\left[\left(t-t'\right)x\right]. \label{eq:i2}
\end{align}
\es

Consider first the integral $\mathcal I_2$ in Eq.~\eqref{eq:i2}, which was given through
\be
\mathcal I_2=\frac{1}{2}\int_0^\infty\df x\,\sin\left(\frac{\varrho}{2}\sqrt{1+x^2}\right)\sin\left[\left(t-t'\right)x\right].
\ee
Introduce $a=\tfrac12(t-t')$ and $b=\tfrac12\varrho$ so that
\be
\mathcal I_2=\frac{1}{2}\int_0^\infty\df x\,\sin\left(b\sqrt{1+x^2}\right)\sin\left(ax\right).
\ee
A known integral \citep[p.~140]{obe57:fou} is given as
\be\label{eq:tmp3}
F(a,b,c)=\frac{1}{2}\int_0^\infty\frac{\df x}{\sqrt{c+x^2}}\,\sin\left(ax\right)\cos\left(b\sqrt{c+x^2}\right).
\ee
Assuming $c=1$, it is easy to see that
\be
-\pd[F]b=\mathcal I_2.
\ee
But the integral $F$ can be rewritten so that
\be
F(a,b,c)=G(a,b,c)-\frac{1}{2}\int_0^{\pi/2}\df t\,\cos\left(bc\cos t\right)\sinh\left(ac\sin t\right),
\ee
where
\be
G(a,b,c)=\frac{\pi}{4}\,I_0\left(c\sqrt{a^2-b^2}\right)
\ee
if $a>b$ and zero otherwise. Here, $I_\nu(\cdot)$ is the modified Bessel function of the first kind of order $\nu$ \citep[see, e.\,g.,][]{ab:math,gr:int}.
By taking the derivative with respect to $b$ and by substituting $x=\sin t$, it can be shown that the remaining integral in Eq.~\eqref{eq:tmp3} is identical to $\mathcal I_1$ and is thus canceled out. The result is
\be
\mathcal I_1+\mathcal I_2=-\pd[G]b=\frac{\pi}{4}\,\frac{bc}{\sqrt{a^2-b^2}}\,I_1\left(c\sqrt{a^2-b^2}\right).
\ee

Collecting terms and re-inserting the parameters used in Eq.~\eqref{eq:i1i2}, the final result can be expressed in closed form as
\be\label{eq:sol2}
G=\frac{2\pi^3}{\varrho}\exp\left[-\frac12\left(t-t'\right)\right]\frac{t-t'}{\sqrt{\left(t-t'\right)^2-\varrho^2}}\,I_1\left(\frac{1}{2}\sqrt{\left(t-t'\right)^2-\varrho^2}\right). 
\ee

As already discussed in Sec.~\ref{norm}, the particle number is conserved only if the corresponding integral over Green's function is independent of $t$. By integrating Eq.~\eqref{eq:sol2} over the spatial coordinates, the result for the particle number is
\begin{align}
N(t)&=\int\df^3r\;G(\f r,t)=4\pi\int_0^\infty\df r\;r^2G\nonumber\\
&=4\pi^3t\exp\left[-\tfrac{1}{2}\left(t-t'\right)\right]\left\{I_0\left(\tfrac{1}{2}\left(t-t'\right)\right)-1\right\},
\end{align}
which is not only variable in time---as was already the case with Eq.~\eqref{eq:nrm}---but increases linearly for large times. Equation~\eqref{eq:sol2} is therefore clearly unsuited as a solution describing the evolution of a particle ensemble.



\end{document}